# Lunar Volatiles and Solar System Science

A White Paper Submitted To
The Decadal Survey in Planetary Science and Astrobiology 2023-2032

Submitted By:

Parvathy Prem[1]*, Ákos Kereszturi[2], Ariel N. Deutsch[3], Charles A. Hibbitts[1], Carl A. Schmidt[4], Cesare Grava[5], Casey I. Honniball[6], Craig J. Hardgrove[7], Carlé M. Pieters[3], David B. Goldstein[8], Donald C. Barker[9], Debra H. Needham[10], Dana M. Hurley[1], Erwan Mazarico[6], Gerardo Dominguez[11], G. Wesley Patterson[1], Georgiana Y. Kramer[12], Julie Brisset[13], Jeffrey J. Gillis-Davis[14], Julie L. Mitchell[15], Jamey R. Szalay[16], Jasper S. Halekas[17], James T. Keane[18], James W. Head[3], Kathleen E. Mandt[1], Katharine L. Robinson[19], Kristen M. Luchsinger[20], Lizeth O. Magaña[5], Matthew A. Siegler[12], Margaret E. Landis[21], Michael J. Poston[5], Noah E. Petro[6], Paul G. Lucey[22], Rosemary M. Killen[6], Shuai Li[22], Shyama Narendranath[23], Shashwat Shukla[24], Thomas J. Barrett[25], Timothy J. Stubbs[6], Thomas M. Orlando[26], and William M. Farrell[6].

[1]Johns Hopkins University Applied Physics Laboratory;
[2]Konkoly Astronomical Institute, Hungary; [3]Brown University; [4]Boston University;
[5]Southwest Research Institute; [6]NASA Goddard Space Flight Center; [7]Arizona State University;
[8]University of Texas at Austin; [9]MAXD, Inc.; [10]NASA Marshall Space Flight Center;
[11]California State University San Marcos; [12]Planetary Science Institute;
[13]University of Central Florida; [14]Washington University in St. Louis;
[15]NASA Johnson Space Center; [16]Princeton University; [17]University of Iowa;
[18]California Institute of Technology, NASA Jet Propulsion Laboratory;
[19]Lunar and Planetary Institute, USRA; [20]New Mexico State University;
[21]University of Colorado Boulder; [22]University of Hawai‘i at Mānoa;
[23]Indian Space Research Organisation; [24]University of Twente, the Netherlands;
[25]The Open University, UK; [26]Georgia Institute of Technology.

*Contact: Parvathy.Prem@jhuapl.edu; 512-669-9612.

Author Contributions: PP wrote the initial draft of this white paper and integrated text written by CG, CIH, DBG, GYK, JJGD, JLM, JSH, KLR, MAS, MEL, PGL, SS, and WMF. AK, AND, CAH, CAS, CJH, CMP, DCB, DHN, DMH, EM, GD, GWP, JB, JRS, JTK, JWH, KEM, KML, LOM, MJP, NEP, RMK, SL, SN, TJB, TJS, and TMO participated in discussions or reviewed drafts of this paper, and provided many ideas and edits that were incorporated into the paper.

## Introduction

As the writing of the 2013–22 Decadal Survey drew to a close, a revolution in our understanding of the lunar volatile system had just begun. Chandrayaan-1, Deep Impact, and Cassini had all returned evidence for the unexpected, widespread presence of hydroxyl or water on the illuminated lunar surface [1]. Laboratory measurements using powerful modern analytical techniques confirmed that some lunar samples contain water indigenous to the Moon, challenging the decades-old idea of an anhydrous lunar interior [2]. Meanwhile, the LCROSS mission had detected an intriguing variety of volatiles in a permanently shadowed region of Cabeus crater [3], and LRO had just begun its multi-wavelength observations of the lunar surface and exosphere. From 2013–2014, the LADEE mission monitored the lunar exosphere from equatorial orbit, while the two ARTEMIS heliophysics probes continued to monitor exospheric ions from a more distant vantage point. Data analysis, sample analysis, lab experiments and numerical modeling continue to yield new insights — and new questions.

Despite these advances, many fundamental questions regarding the lunar volatile system remain unanswered, and will continue to remain so unless concerted steps are taken over the next decade. This decade also brings the possibility that the lunar environment may be irrevocably altered by increased activity, making certain polar and exospheric measurements time-critical.

Understanding the origin and evolution of the lunar volatile system is not only compelling lunar science, but also fundamental Solar System science:

- Understanding the Moon's primordial water content is critical in understanding the **formation and early evolution of the Earth-Moon system and the inner Solar System**. The volatile history of the lunar interior is recorded in returned lunar samples, pyroclastic deposits, and polar regions that may preserve traces of ancient volcanic outgassing.
- Permanently shadowed regions (PSRs) near the lunar poles hold a unique record of the **delivery of water** to the inner Solar System over the past several billion years, and bear witness to the processes that have shaped our local space environment.
- Understanding the processes that control global surface hydration on the Moon is critical to understanding observations of hydration on other airless bodies, and thereby the **origin and distribution of water** throughout the Solar System [4]. **Comparative planetology** of ice on airless bodies may yield insights into the evolution of such bodies across the Solar System.
- PSRs are among the coldest places in the Solar System, harboring a variety of volatile species, including water, hydrocarbons, and nitrogen- and sulfur-bearing compounds. The **lunar polar microenvironment** is a **natural laboratory** in which to study abiotic/prebiotic chemistry and other surface processes that may be active elsewhere in the Solar System.
- The lunar exosphere remains our closest example of the **most common class of atmosphere in the Solar System** — a surface boundary exosphere. However, the density of the lunar atmosphere may have varied dramatically over time. Understanding the evolution of the lunar atmosphere has implications for our fundamental understanding of **how atmospheres rise and fall**, and the **behavior of rarefied atmospheres across the Solar System**, from the Moon and Mercury to asteroids, outer Solar System satellites, and beyond.

This white paper briefly summarizes advances in our understanding of lunar volatiles over the past decade, identifies outstanding questions for the next decade of planetary science and astrobiology, and discusses key steps required to address these questions.

## The Lunar Volatile Revolution

Until the late 2000s, the Moon was considered dry, devoid of water and most other volatiles, with the potential exception of cold PSRs near the lunar poles [5]. Returned lunar samples contained no hydrous minerals or weathering products indicating exposure to water, while bulk measurements of Apollo basalts showed they were essentially dry [6]. Over the past decade, a range of laboratory and remote sensing techniques have found evidence for the presence of water and other volatiles not only in some PSRs, but also in the lunar interior, and on the sunlit lunar surface.

Since 2008, laboratory studies of Apollo samples using improved analytical techniques have found traces of water in volcanic glasses and olivine-hosted melt inclusions [2], as well as the mineral apatite [7], overturning the previously accepted view of an anhydrous lunar interior. These measurements further indicate that the isotopic composition and concentration of water may vary within the lunar interior, with implications for the formation and evolution of the Earth-Moon system and the inner Solar System [8]. More recently, indigenous hydration has also been detected remotely in pyroclastic deposits [9] and material excavated from depth [10].

Before the arrival of LRO/LCROSS at the Moon in 2009, Lunar Prospector Neutron Spectrometer (LPNS) measurements had found evidence for enhanced subsurface hydrogen at the lunar poles, generally coincident with cold traps [11]. Potential detections of water ice near the lunar south pole by the Clementine bistatic radar experiment had not held up to further scrutiny [12]. Since then, the LRO LEND instrument [13] and further analysis of the LPNS dataset [14] have advanced our understanding of the distribution of subsurface hydrogen, but critically, it remains undetermined whether this hydrogen is mainly in the form of water ice or chemically bound minerals.

There is evidence for the presence of surficial water within PSRs from UV observations [15], laser reflectance measurements [16], and near-infrared spectra [17], which probe the upper nanometers to millimeters of the lunar surface. Positive detections of water coincide at some locations, but not at others. Intriguingly, LRO laser reflectance data hint at the presence of other volatiles, potentially sulfur or organics [16]. The most definitive evidence for polar volatiles comes from LCROSS, which found multiple volatile species ($H_2O$, $H_2S$, $NH_3$, $SO_2$, $CO_2$, and others) in Cabeus crater [3]. Recent bistatic radar measurements from LRO and Earth-based observatories also suggest the presence of buried ice at Cabeus, but these observations remain challenging to interpret [18].

Beyond the poles, near-infrared datasets from three separate missions show near-global absorption features consistent with the presence of $H_2O$ or OH on the sunlit surface of the Moon [1]. However, IR datasets require careful correction for the effects of thermal emission at these wavelengths, and different correction techniques have led to differing estimates of water abundance and variability [19]. Recent LRO UV observations suggest sub-monolayer levels of diurnally migrating $H_2O$, but do not rule out the presence of OH [20]. Most recently, $H_2O$ has been detected using the airborne SOFIA observatory to view the Moon at wavelengths that characterize $H_2O$ less ambiguously [21].

Placing remote sensing observations into context requires an understanding of volatile sources, sinks, surface chemistry, physical mobility, sequestration, and loss processes. Critical information in this respect has come from sample analysis, laboratory experiments, and numerical modeling. Detailed thermal modeling indicates that polar cold traps have varied in extent over the past several billion years, and that the observed distribution of sub-surface hydrogen today is more consistent with deposition in a past thermal environment [22]. Meanwhile, sample analyses, coupled with estimates of mare basalt volume, indicate that volcanic outgassing could have released *more than enough* water to account for all the hydrogen presently observed at the lunar poles [23]. However, the critical question of how efficiently water can migrate to the poles depends on how strongly molecules bind to the lunar regolith — a subject of ongoing lab investigations [24]. These are just a few illustrative examples of recent studies.

Elsewhere in the Solar System, the confirmation of relatively thick, pure deposits of water ice (as well as dark, likely organic, volatiles) at Mercury's poles raises the intriguing question of how two rocky bodies in the inner Solar System came to be so different, and whether differences in sources, processes, or timing are responsible [25]. Scientific opportunities afforded by Mercury's polar deposits are summarized in the white paper by Deutsch, Chabot, et al., and similar opportunities exist at the lunar poles. Meanwhile, although Ceres differs from the Moon in the presence of a buried ice table, PSRs on Ceres have also been found to harbor water ice [26].

The lunar surface and exosphere are intimately connected. Each volatile species tells an important story about the origin and evolution of surface boundary exospheres throughout the Solar System. Prior to the last decade, we had obtained direct detections of only a handful of neutral species in the lunar exosphere, and only upper limits on most others. Since then, LADEE and LRO have constrained the abundance and variability of helium and argon, and revealed previously undetected species, including neon [27], methane [28], and molecular hydrogen [29]. Some exospheric species have been first and/or solely detected in charged form. Observations of solar wind-accelerated "pickup ions" in the further reaches of the lunar exosphere by the ARTEMIS heliophysics mission have played an important role in understanding exospheric origin and dynamics [30].

Many studies of exospheric volatiles have been motivated by the desire to understand the fate of the solar wind upon encountering the surface of an airless body — an interaction that occurs across the Solar System. Recent data suggest that 10–50% of incoming solar wind $H^+$ may be converted to $H_2$ [31], and that 25–76% of solar wind $C^+$ may escape the Moon as $CH_4$ [28]. The significant ranges of uncertainty in these estimates reflect fundamental gaps in our understanding of the solar wind hydrogen and carbon cycles.

Radiogenic argon ($^{40}Ar$) was among the first species to be detected in the lunar exosphere, but its origin and behavior have been revisited this decade in light of density variations observed by LADEE [27]. These variations have been attributed to outgassing from the Oceanus Procellarum region [32], or seasonal migration of $^{40}Ar$ between north and south polar cold traps [33]. The alkali metals Na and K are trace species, yet observable from Earth, and originate primarily from the lunar surface, motivating ongoing work that aims to correlate surface and exospheric composition [34]. Curiously, water, the best-studied volatile on the lunar surface, remains elusive in the exosphere — except for brief bursts of vapor coincident with micrometeoroid showers. LADEE observations have led to the recent hypothesis that micrometeoroids may liberate $OH/H_2O$ from a global sub-surface reservoir [35], a mechanism that is supported by laboratory experiments [36].

Scientific exploration of the hydrous Moon is just beginning, and the lunar volatile revolution is far from complete.

## Science Questions for the Next Decade

The lunar volatile system has fundamental connections to all three of the cross-cutting themes identified in the 2013–22 Decadal Survey: understanding Solar System beginnings, searching for the requirements for life, and revealing planetary processes through time. The Scientific Context for Exploration of the Moon (SCEM) was comprehensively described in a 2007 report of the same name [37]. Many fundamental questions about lunar volatiles (SCEM Science Concepts 4 and 8) remain unanswered, as identified in the 2018 Advancing Science of the Moon (ASM) report [38]. The most critical science questions for the next decade include the following:

**a. What is the composition (elemental and isotopic), concentration, distribution (lateral and vertical), and physical form of polar volatiles?**

This is a fundamental question that remains largely unanswered, and is crucial to addressing many of the questions that follow. Knowing the current state of the polar microenvironment is critical for testing existing hypotheses regarding origins and processes, and developing new ones. The distribution and physical form of surface and sub-surface hydration and hydrogenation remain to be definitively characterized, and we have scarcely begun to take inventory of other volatile species or isotopic compositions, which may serve as important tracers of origins and processes.

**b. What are the relative contributions of impacts, volcanism and solar wind to the lunar polar volatile inventory, and how have these sources varied over time?**

We do not currently understand whether the water present at the lunar poles is ancient (originating predominantly from water-rich impactors or volcanic outgassing) or relatively modern (originating predominantly from ongoing solar wind and micrometeoroid bombardment of the lunar surface). We know that impact flux and volcanic activity have varied dramatically through time, but the evolution of polar cold traps (linked to both crater age [39] and orbital evolution [40]) is much less well-understood. We also do not understand the extent to which surface and sub-surface reservoirs may be connected or independent. Origin and timing are critical aspects of the unique geological record written at the lunar poles, as well as important clues to solving the puzzle of why the polar regions of the Moon and Mercury are so different.

**c. What are the transport, retention, alteration, and loss processes that operate on volatiles in the lunar polar environment?**

Understanding the processes that preserve, alter, and destroy volatiles on airless bodies is critical to deciphering the polar volatile record, including why volatiles appear to be so heterogeneously distributed (between the lunar north and south poles, as well as between and within individual craters), and whether the Moon is in a state of net volatile accumulation, loss, or balance. Impact gardening [41], plasma sputtering [42], and temperature-driven migration [43] of volatiles are all thought to be important, but have never been studied in situ. Recent models suggest that high-inclination meteoroids may be particularly active in redistributing polar material [44]. LCROSS results hinted at the possibility that surface chemistry on cold grains could alter the composition of cold-trapped volatiles [3], and the recent detection of hematite at high latitudes [45] suggests that alteration minerals may be more prevalent than previously recognized. Chemistry that may be driven by energy from radiation and impacts at the lunar poles remains to be studied in detail, including the potential synthesis of organics [46] and clathrates [47], or silicate alteration [48].

### d. What is the distribution and physical form of lunar surface volatiles beyond the poles?

It remains to be determined what fraction of the global hydration signature discovered on the sunlit lunar surface at the beginning of the decade is attributable to hydroxyl and/or molecular water, and whether this signature varies spatially and temporally. The formation of volatiles (not only OH or $H_2O$, but also $H_2$, $CH_4$ and other species) by solar wind bombardment may be an integral part of space weathering on all airless bodies. Subsurface hydration beyond the poles has been inferred from LADEE observations [35], but remains to be established. In addition to the global hydration feature, detailed characterization of the volatile enhancements associated with pyroclastic deposits and craters could provide unique windows into the evolution of the lunar interior, complementing results from analysis of returned samples.

### e. How does the contemporary lunar volatile cycle operate?

There are significant gaps in our understanding of the contemporary volatile cycle: the extent to which different volatile species migrate across the lunar surface (including whether such migration occurs at all), how ions and neutrals exchange energy with the surface, and the sensitivity of those interactions to temperature and composition. Our knowledge of the composition and variability (spatial and temporal) of surface and exospheric volatiles is fundamentally incomplete. We are just beginning to understand how the lunar exosphere responds to solar wind [49] and micrometeoroid bombardment [35], as well as temporal variations in sources and sinks. Spacecraft operations and human presence will almost inevitably be active volatile release experiments that will *at least* temporarily perturb the tenuous lunar exosphere, presenting both a scientific opportunity and an operational need to understand how the contemporary volatile cycle operates.

### f. How has the lunar atmosphere changed over time, and how is this history preserved?

The longevity of potential past lunar atmospheres — from an outgassed, primordial atmosphere [50], to more recent transient atmospheres generated by volatile-rich impacts [51] or volcanic eruptions [23] — has fundamental implications for our understanding of planetary atmospheric evolution. Important characteristics of the lunar environment, such as polar temperatures [40] and magnetic field strength [52], have also varied over time. Understanding how and if traces of past atmospheres are preserved remain important outstanding questions with implications for the origin of polar volatiles, the evolution of the Earth-Moon system, and other planetary bodies. Reconstructing the origin and fate of transient atmospheres requires theoretical work, sample/data analysis, and characterization of the stratigraphy and composition of volatiles at the lunar poles.

### g. Is lunar water a viable resource for Solar System exploration?

The past decade has seen widespread interest in the potential usage of lunar water as a resource. Determining resource viability requires an assessment of the abundance, distribution, physical form, and the physical and technological accessibility of water in polar and non-polar regions. Knowledge of other volatiles that may be present and the renewability (or lack thereof) of lunar water, are also important considerations in planning in situ resource utilization. We highlight resource viability as a separate science question due to the wide interest it draws beyond the planetary science community, and the broader implications for how we explore the Solar System. Similar to planetary defense, resource viability is a question of both fundamental scientific as well as immediate practical interest.

## The Next Decade

The outstanding science questions above could be addressed within the next decade through a range of mission architectures, accompanied by support for research and analysis, technology development, and international partnerships. LRO, currently in its fourth extended mission, was developed at a time when our understanding of and questions regarding the lunar volatile system were much simpler. Similarly, Chandrayaan-1's Moon Mineralogy Mapper instrument, which returned data that continue to play a pivotal role in our understanding of surface hydration, was not in fact designed to characterize surficial volatiles. Indeed, the 2018 ASM report identified the lunar volatile cycle as a 'new concept' that has emerged and grown in importance over the past decade. Remarkable advances have been gained by stretching existing orbital datasets to their limits, but fundamental gaps in our understanding of the lunar volatile system remain.

Some of the measurements that would contribute to addressing these gaps are time-critical. Both the polar microenvironment and the tenuous lunar exosphere are susceptible to disturbances caused by robotic and human activity, and unless characterized carefully over the next decade, could be irretrievably lost as a scientific resource as exploration of the lunar surface accelerates.

**Orbital and ground-based observations:** Orbital missions remain a powerful way to gain an understanding of the global distribution of volatiles (water and others), the associated geological context, and a view of the volatile system as an integrated whole. Long-term missions enable extended coverage and improved signal-to-noise, as well as a means to observe the spatial and temporal response of the volatile system to changes in the lunar environment: the ebb and flow of seasonal shadow at the poles, the liberation of volatiles during micrometeoroid showers, seismic activity, and spacecraft landings, and changes in solar wind flux during magnetotail passages.

SmallSat and CubeSat missions (e.g. [53]) can complement Discovery-class and larger missions through the ability to acquire low-cost measurements that respond to evolving science questions, and should continue to be supported. However, holistic characterization of the volatile system may require longer-lived spacecraft carrying a broader range of instrumentation, as discussed in the white paper by Lucey et al. Earth-based observations from ground-based or airborne platforms can act as pathfinders for orbital remote sensing techniques, complementing spacecraft observations by providing access to unique viewing geometries and different wavelengths.

**Surface science:** Volatiles science at the lunar surface is driven by the need to characterize the distribution of volatiles on scales unresolvable from orbit, to characterize surface processes in situ, and to directly access the sub-surface. In situ measurements at unique locations (e.g., lunar swirls, where magnetic fields modulate the interaction of the solar wind at the surface, or pyroclastic deposits, where indigenous water may be accessible) may be especially informative.

NASA's Commercial Lunar Payload Services program plans to deliver several instruments to the lunar surface over the next decade, some of which will make volatiles-related measurements. There is also significant value in a long-lived, global network of landed platforms that could measure diurnal and seasonal variations in surface and exospheric volatile abundance, detect localized sources, monitor the migration of volatiles to polar cold traps, and characterize the sensitivity of gas-surface interactions to composition and thermophysical properties. Mobile architectures can greatly enhance science return, and may be required to achieve certain science goals. In particular, the patchy distribution of volatiles and the variability of the polar environment (e.g., temperature and solar wind flux) drive a need for mobility for polar missions, as discussed in the white paper by Hurley et al. The development of systems that can survive and operate on the lunar surface for long durations should be supported, including the development of electronic systems that can operate in thermal extremes and power systems that can sustain operations during the lunar night.

When planning surface operations it is crucial to bear in mind that the polar microenvironment has evolved over billions of years, and that our exploration of this environment will inevitably alter (temporarily or permanently) its thermophysical character, and very likely introduce new volatiles to the system. Understanding and planning for the impact of exploration on the lunar environment is critical to carrying out meaningful and responsible science.

**Sample collection and return:** Despite continued improvements in analytical instrumentation, there remain critical measurements that cannot be performed outside a lab. Sample return would enable a greater range of measurements, with higher sensitivity and resolution, and allow for the long-term preservation of samples for future scientific study. Technology development over the next decade should address the unique challenges associated with the return of volatile samples from Solar System bodies, which are discussed in the white paper by Milam et al. To enable sample collection by humans on the Moon, technology development should include the development of tools that allow for in situ analysis and sample acquisition in hitherto unexplored environments.

**Research and analysis:** Data and sample analysis, laboratory work, and modeling have all played critical roles in furthering our understanding of lunar volatiles over the past decade. Continued investment in these areas is vital. Support for data analysis has demonstrably magnified the science return from existing datasets. Recent advances in analytical techniques continue to enable new science from old samples. Modeling is often instrumental in reconstructing the past, and to predicting and interpreting present-day observables. Laboratory studies can provide crucial information on lunar environmental processes, including the mechanisms of water production on the lunar surface, desorption kinetics, and surface chemistry. The Solar System Exploration Research Virtual Institute (SSERVI), which provides five years of stable funding for a large team, has been an effective mechanism for advancing the science of lunar volatiles over the past decade.

Several space agencies currently have plans for missions that would investigate the lunar volatile system. Major recent advances have resulted from the participation of US scientists in international missions, and vice versa. Support for such collaborations should continue through participating scientist programs and other measures to broaden participation in US and international missions.